\documentclass[runningheads,a4paper]{llncs}

\usepackage{amssymb}
\usepackage{graphicx}
\usepackage{longtable}

\usepackage{latexsym} %Box and Diamond
\usepackage{listings} 
\usepackage{todonotes}
\usepackage{makecell}
\usepackage{tikz}
\usetikzlibrary{shapes,positioning, arrows, calc, fit}

\usepackage{url}

\newcommand{\keywords}[1]{\par\addvspace\baselineskip
\noindent\keywordname\enspace\ignorespaces#1}

\newcommand{\moise}{\ensuremath{\mathcal{M}\textsc{oise}^+}}
\newcommand{\aorta}{\textsf{AORTA}}

\begin{document}
\renewcommand{\thelstlisting}{\arabic{lstlisting}}

\title{Model Checking AORTA:\\Verification of Organization-Aware Agents}

\author{Andreas Schmidt Jensen}

\institute{
Technical University of Denmark, Kongens Lyngby, Denmark \\
\path|ascje@dtu.dk| \\
}

\maketitle
\thispagestyle{plain} \pagestyle{plain}

\begin{abstract}
As agent systems grow larger and more complex, there is an increasing need to formally verify them. Furthermore, it is often suggested that complex systems can be regulated using organizational models, imposing constraints on the agents in the systems. Agents that can understand the organizational model and constraints in a system is said to be \emph{organization-aware}. This paper is concerned with verification of organization-aware agents. We show how agents using \aorta, a framework for making agents organization-aware, can be formally verified using an extended version of the Agent Java PathFinder (AJPF), a model checking system designed specifically for agent programming languages. We integrate \aorta\ with the Agent Infrastructure Layer (AIL), which is an intermediate layer on top of which APLs can be implemented, and use our extension of AJPF to verify a system of agents aiming to write a paper together by using an organization for coordination.
\end{abstract}

\section{Introduction}
In many of the areas where multi-agent systems (MASs) are used, there is a need for dependability and security. Therefore, it is increasingly necessary to consider formal verification of such systems \cite{Bordini+2008}. Furthermore, we have seen an increase of interest in the area of organization-oriented MAS, i.e. systems in which agents have to consider organizational constraints. The motivation for organizational MASs is the increasing complexity of \emph{heterogeneous agents} in \emph{open systems}. The owner of an open system cannot in general assume much about agents entering the system, and it is therefore important to be able to regulate their behavior to ensure that it is within the acceptable boundaries of the system. 

Organizational models (e.g. \moise\ \cite{Hubner+2002}) are designed to describe what is expected of agents in the system without taking the individual agents and their implementation into account. This is done using the notion of roles: an agent can a enact roles, giving it certain responsibilities (the objectives of the role) while providing certain capabilities (access to objects in the system, access to groups of other agents, etc.). Furthermore, organizational models often have a normative aspect: behavior, or states of affair, that is expected of the agents in the system, but is not directly enforced. That is, the agents are free to violate the norms of a system, but they should then expect to be punished. Organizational models thus provide a way for the designer of a system to explain to the agents entering the system, what is expected of them. 

However, if agents are expected to fulfill the system's expectations of them, they need a way to understand the organizational model of that system. Agents that are able to do this are \emph{organization-aware} \cite{Boissier+2013}. Organization-aware agents will naturally tend to be more complex than their ``unaware'' counterparts; even though programming them may be easier, since certain aspects may be automated (task allocation, coordination, etc.), the reasoning cycle of the agents will include more steps. This makes it even harder to convince ourselves that our implementation is correct. 

Since agent-oriented programming (AOP) differs from the well-known object-oriented programming (OOP), the verification techniques from OOP must be extended to capture the agent metaphor. That is, since agents are autonomous and their behavior is based on beliefs and intentions, we need to be able to not only check \emph{what} the agent does (similar to verification in OOP), but also \emph{why} it did so.

The principles of model checking as defined in \cite{Baier+2008} ``\emph{is an automated technique that, given a finite-state model of a system and a formal property, systematically checks whether this property holds for (a given state in) that model}''. Model checking agent programming languages (APLs) can thus be reduced to translating the system into a finite-state model in which we can prove certain properties. Since agents are usually enriched with mental attitudes such as beliefs, goals and intentions, model checking APLs is only interesting, if we can verify properties about these mental attitudes. For example, it is possible to check whether agents in a system only intend to achieve goal states by checking the temporal formula $\Box(\textbf{I}(\emph{ag}, \phi) \rightarrow \textbf{G}(\emph{ag}, \phi))$. Here, \emph{ag} refers to an agent, $\phi$ is a state and \textbf{I} and \textbf{G} are modal operators referring to intentions and goals, respectively. Quite some work has been done to make it possible to perform model checking on existing APLs, and for example, the Agent Java PathFinder project \cite{Dennis+2011} is an example of a practical system in which model checking is feasible.

In this paper, we present an extension to AJPF, which makes it possible to perform verification of organization-aware agents. We use \aorta\ \cite{Jensen+2014a} to make agents organization-aware, and extend the specification language to incorporate modalities about organizational information. Our contribution is two-fold: first, we integrate \aorta\ with the Agent Infrastructure Layer (AIL), which is an intermediate layer on top of which APLs can be implemented. By integrating \aorta\ with AIL, we enable verification of organization-aware agents from potentially any APL with a clear semantics. Second, we use the integration to verify properties in a system of agents working together to write a scientific paper. 

The rest of the paper is organized as follows. First, we provide a brief description of the \aorta\ framework in section \ref{sec:aorta}. We then describe the AJPF system, which is used to translate MASs into finite-state models on which model checking can be done (section \ref{sec:mcapl}). In section \ref{sec:mco} we present our extension to AJPF, making it possible to perform model checking on organization-aware agents. We evaluate the system in section \ref{sec:eval} and we conclude the paper in section \ref{sec:conc}.

\section{The AORTA framework}\label{sec:aorta}
\aorta\ \cite{Jensen+2014a} is an organizational reasoning component that can be integrated into an agent's reasoning mechanism, allowing it to reason about (and act upon) regulations specified by an organizational model using reasoning rules. That is, the organization is preexisting and independent from the agent and the component is agent-centered, focusing on letting the agent reason about the organization. By separating the organization from the agent, the architecture of the agent is independent from the organizational model, and the agent is free to decide on how to use \aorta\ in its reasoning. The separation is achieved by basing the component on reasoning rules using an organizational metamodel, designed to support different organizational models. A prototype of \aorta\ has been implemented in Java\footnote{Java was chosen since many existing agent platforms are built in Java.} \cite{Jensen+2014b}, designed such that it can provide organizational reasoning capabilities to agents implemented in existing APLs. 

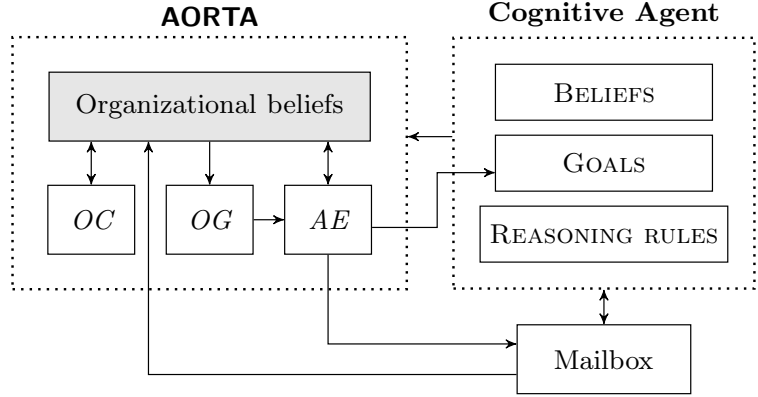
\begin{figure}[t]
  \centering      
    \resizebox{4in}{!}{
      %!TEX root = modelchecking.tex
\begin{tikzpicture}[>=stealth',->]

\node [rectangle, draw, fill=gray!20,
	minimum height=0.8cm,
	minimum width=3.7cm
] (organization) {Organizational beliefs};

\node [below=0.5cm of organization,
	rectangle, draw, 
	minimum width=1cm,
	minimum height=0.8cm,
	font=\em
] (opt) {
	OG
};

\node [right=0.35cm of opt,
	rectangle, draw, 
	minimum width=1cm,
	minimum height=0.8cm,
	font=\em
] (act)  {
	AE
};

\node [left=0.35cm of opt,
	rectangle, draw, 
	minimum width=1cm,
	minimum height=0.8cm,
	font=\em
] (obl) {
	OC
};

\node [fit=(obl)(opt)(organization)(act),label=\textbf{\aorta},draw,dotted,thick,rectangle,inner sep=0.4cm] (orgcomponent) {\hspace*{\fill}~}; % to avoid underfull hbox

\draw[<->] (obl.north) -- (obl.north |- organization.south);
\draw[<->] (act.north) -- (act.north |- organization.south);
\draw (organization) -- (opt);
\draw (opt) -> (act);

\node [
	rectangle,draw,
	minimum width=2.5cm,
	minimum height=0.65cm,
	right=1cm of orgcomponent
] (goals) {\textsc{Goals}};
\node [
	rectangle,draw,
	minimum width=2.5cm,
	minimum height=0.65cm,
	above=0.16cm of goals
] (bel) {\textsc{Beliefs}};

\node [
	rectangle,draw,
	minimum width=2.5cm,
	minimum height=0.65cm,
	below=0.16cm of goals
] (rules) {\textsc{Reasoning rules}};

\node [inner sep=0.3cm,fit=(bel)(goals)(rules),draw,dotted,thick,rectangle,label=\textbf{Cognitive Agent}] (cognitive) {\hspace*{\fill}~};

\draw (cognitive.170) -- (cognitive.170 -| orgcomponent.east);
\draw (act.350) -- +(0.75,0) |- (goals.185);

\node [
	rectangle, draw,
	minimum width=2cm,
	minimum height=0.8cm,
	below=0.4cm of cognitive
] (mbox) {Mailbox};

\draw (mbox.190) -| (organization.210);
\draw[<->] (mbox) -- (cognitive);
\draw (act) |- (mbox.170);

\end{tikzpicture}
    }
    \caption{The \aorta\ component. The arrows indicate flow of information. Obligations and options are generated from the organizational beliefs, and actions are based on the generated options.}
  \label{fig:component}
\end{figure}

%\aorta\ provides organizational reasoning capabilities to agents, allowing the agents to reason about organizational matters. 
In this paper, we use an extended version of the \aorta\ architecture described in \cite{Jensen+2014b}: options are generated automatically, action deliberation and coordination are merged into a single phase, and we incorporate obligations. Organizational reasoning in \aorta\ is then divided into three phases: \emph{obligation check} (OC), \emph{option generation} (OG) and \emph{action execution} (AE). The OC-phase uses the agent's mental state and organizational state to determine if obligations are activated, satisfied or violated, and updates the organizational state accordingly. The OG-phase uses the organizational specification to generate possible organizational options. The agent considers these options in the AE-phase using reasoning rules, which can alter the organizational state and the agent's intentions, or send messages to other agents. The component is shown in figure \ref{fig:component}. We assume is connected to a \emph{cognitive} agent, i.e., agents with mental attitudes (such as beliefs and goals) and practical reasoning rules.

\paragraph{Checking obligations:} 
In the OC-phase, \aorta\ uses the agent's state to determine for each obligation if it should change to a new state. We use conditional obligations with a deadline, so an obligation can change state in several different situations. If the condition for activating an obligation has happened, the component activates the obligation by updating the organizational state. Similarly, it checks whether the obligation has been satisfied (the objective is completed) or violated (the deadline was reached before the objective was completed). 

\paragraph{Option generation:}
In the OG-phase, \aorta\ uses the mental state of the agent and the organizational state to consider what the agent can do regarding the organization. The following organizational aspects are considered in the OG-phase:
\begin{description}
  \item[\textbf{Role enactment:}] Roles that are possible to enact given the agent's goals.
  \item[\textbf{Role deactment:}] (Currently enacting) roles that have been fulfilled or are no longer useful.
  \item[\textbf{Obligations:}] States the agent is currently obliged to achieve.
  \item[\textbf{Delegation:}] Objectives that can be delegated based on a dependency relation.
  \item[\textbf{Information:}] Obtained information that other agents will benefit from knowing.
\end{description}
The options that are generated in this phase are then available to act upon in the AE-phase.

\paragraph{Action execution:}
The AE-phase uses reasoning rules to decide how to react on a given option in a given context. The AE-phase selects at most one option to act upon. The reasoning is based on rules of the form\footnote{Inspired by the plan syntax of AgentSpeak(L) \cite{Rao+1996}.}
$$
\emph{option}~:~\emph{context}\rightarrow\emph{action}
$$
where \emph{option} is a previously generated option, \emph{context} is a state description that should hold for an action to be applicable, and \emph{action} is the action to be executed. 

The agent has actions available to enact or deact a role, commit to complete or drop an objective, and send messages. This corresponds to the options that can be generated in the previous phase. 

\subsection{The AORTA organizational metamodel}
Reasoning in \aorta\ uses an organizational metamodel, which is based on roles, objectives and obligations, as these concepts are commonly used in existing organizational models (e.g. \moise).

\begin{definition}[Organizational metamodel]
The organizational metamodel of \emph{\aorta} is defined by the following predicates:

\begin{center}\footnotesize
\begin{tabular}{lp{7cm}}
\emph{\textsf{role}}($\mathit{Role}, \mathit{Objs}$) & $\mathit{Role}$ is the role name, and $\mathit{Objs}$ is a set of objectives. \\[3pt]
\emph{\textsf{obj}}($\mathit{Obj}$, $\mathit{SubObjs}$) & $\mathit{Obj}$ is the name of an objective, and $\mathit{SubObjs}$ is a set of sub-objectives. \\[3pt]
\emph{\textsf{dep}}($\mathit{Role}_1, \mathit{Role}_2, \mathit{Obj}$)~~~& Role $\mathit{Role}_1$ depends on role $\mathit{Role}_2$ for completion of objective $\mathit{Obj}$. \\[3pt]
\emph{\textsf{rea}}($\mathit{Ag}, \mathit{Role}$) & Agent $\mathit{Ag}$ enacts role $\mathit{Role}$. \\[3pt]
\emph{\textsf{cond}}(\textit{Role}, \textit{Obj}, \textit{Deadline}, \textit{Cond}) & A conditional obligation for role \textit{Role} to complete \textit{Obj} before \textit{Deadline} when \textit{Cond} holds. \\[3pt]
\emph{\textsf{obl}}(\textit{Ag}, \textit{Role}, \textit{Obj}, \textit{Deadline}) & An obligation for agent \textit{Ag} playing role \textit{Role} to complete \textit{Obj} before \textit{Deadline}. \\[3pt]
\emph{\textsf{viol}}(\textit{Ag}, \textit{Role}, \textit{Obj}) & Agent \textit{Ag} playing role \textit{Role} has violated the obligation to complete \textit{Obj}.
\end{tabular}
\end{center}
\end{definition}

A role is defined only by its name and its \emph{main} objectives. Sub-objectives of an objective are specified using \textsf{obj}-predicates. We distinguish between the different states of obligations by using different predicates. For example, a conditional obligation is represented by the predicate \textsf{cond}(\emph{borrower}, \emph{return}(\emph{Book}), \emph{Deadline}, \emph{borrowed}(\emph{Book})). If an agent $Bob$ enacts the \emph{borrower} role and borrows the book ``1984'', the obligation is activated, which is represented by the predicate \textsf{obl}(\emph{bob}, \emph{borrower}, \emph{return}(\emph{1984}), \emph{Deadline}). A violation of the obligation is represented by the predicate \textsf{viol}(\emph{bob}, \emph{borrower}, \emph{return}(\emph{1984})).

\subsection{Operational semantics of AORTA}
The \aorta\ framework has a well-defined operational semantics, which has been implemented in Java and integrated with \emph{Jason}. We will not go into details with the semantics in this paper, but give the relevant definitions required to understand our integration of \aorta\ with AIL. 

One of the key ideas of \aorta\ is the notion of the organizational knowledge base used by the component for reasoning about options and actions. Furthermore, the component contains an \emph{options base} containing the options generated in the OG-phase.

\begin{definition}[Mental state]
The \emph{\aorta} mental state is based on knowledge bases.  Each knowledge base is based on a predicate language, $L$, with typical formula $\phi$. The agent's belief base and intention base are denoted $\Sigma_{a}$ and $\Gamma_{a}$, respectively. The language of the organization is denoted $L^{org}$, and $L^{org} \subseteq L$, and the option language is denoted $L^{opt}$, and $L^{opt} \subseteq L$. The organizational specification and options are denoted $\Sigma_{o}$ and $\Gamma_{o}$, respectively.
The mental state, \emph{MS}, is then a tuple of knowledge bases:
$$
\textrm{MS} = \langle \Sigma_{a}, \Gamma_{a}, \Sigma_{o}, \Gamma_{o} \rangle,
$$
where $\Sigma_{a},\Gamma_{a} \subseteq L$, $\Sigma_o \subseteq L^{org}$ and $\Gamma_o \subseteq L^{opt}$
\end{definition}

\begin{definition}[Options]
The option language, $L^{opt}$ with typical element $\gamma$ is defined as follows:
$$
\gamma~::=~\emph{\textsf{role}}(R)~|~\emph{\textsf{obj}}(O)~|~\emph{\textsf{send}}(R,\emph{ilf},\phi),
$$
where $R$ is a role identifer, $O$ is an objective, \emph{ilf} is \emph{tell} or \emph{achieve}, and $\phi \in L$ is a message.
\end{definition}

Each of the knowledge bases in the mental state can be queried using \emph{reasoning formulas}.

\begin{definition}[Formulas]
\emph{\aorta} uses reasoning formulas, $L_R$, with typical element $\rho$, which are based on organizational formulas, option formulas, belief formulas and goal formulas:
$$
\rho~::=~\top~|~\emph{\textsf{org}}(\phi)~|~\emph{\textsf{opt}}(\phi)~|~\emph{\textsf{bel}}(\phi)~|~\emph{\textsf{goal}}(\phi)~|~\neg \rho~|~\rho_1\land \rho_2,
$$
where $\phi\in L$.
\end{definition}

Organizational formulas, $\textsf{org}(\phi)$, queries the organizational beliefs, option formulas, $\textsf{opt}(\phi)$, queries the options base, belief formulas, $\textsf{bel}(\phi)$, queries the belief base and goal formulas, $\textsf{goal}(\phi)$, queries the goal base.

\begin{definition}[Semantics of reasoning formulas]
The semantics are based on the agent's mental state, $\textrm{MS} = \langle \Sigma_{a}, \Gamma_{a}, \Sigma_{o}, \Gamma_{o} \rangle$.
$$
\begin{array}{lcl}
\mathit{MS} \models \top &&\\
\mathit{MS} \models \emph{\textsf{bel}}(\phi) &\mathit{iff}& \phi\in\Sigma_{a}  \\
\mathit{MS} \models \emph{\textsf{goal}}(\phi) &\mathit{iff}& \phi\in\Gamma_{a} \\
\mathit{MS} \models \emph{\textsf{org}}(\phi) &\mathit{iff}& \phi\in\Sigma_{o}  \\
\mathit{MS} \models \emph{\textsf{opt}}(\phi) &\mathit{iff}& \phi\in\Gamma_{o} \\
\mathit{MS} \models \neg \rho &\mathit{iff}& \mathit{MS} \not\models \rho \\
\mathit{MS} \models \rho_1 \land \rho_2 &\mathit{iff}& \mathit{MS} \models \rho_1~\mathrm{and}~\mathit{MS} \models \rho_2 
\end{array}
$$
\end{definition}

We define the configuration of an agent with an \aorta\ component as follows:
\begin{definition}[\aorta-agent]
An \emph{\aorta}-agent configuration is defined by the following tuple:
$$
A = \langle \alpha, \emph{MS}, \mathit{AR}, F, \mu \rangle,
$$
where $\alpha$ is the name of the agent, \emph{MS} is the mental state, $\mathit{AR}$ is the agents reasoning rules, $\mathit{AR}\subseteq \mathcal{R}_A$, $F$ is the set of transition functions and $\mu = \langle \mu_{in}, \mu_{out} \rangle$ is the mailbox, contains incoming and outgoing messages.
\end{definition}

The initial configuration consists of a set of initial beliefs and goals, and the organizational specification. The agent has a number of \emph{state transition rules} available, which can be used to change its state. The execution of an entire organizational cycle will check for messages and external changes, apply obligation rules, generate options and execute an action. Then, the reasoning cycle of the agent enriched with the component is executed (e.g. the \emph{Jason} reasoning cycle).

\section{Model Checking Agent Programming Languages}\label{sec:mcapl}

Much of the work done in the area of model checking MASs and agent programming languages has been in the setting of AgentSpeak(L) \cite{Bordini+2004a,Bordini+2006,Bordini+2004b}. While interesting, such approaches are generally hard to extend to other languages without a lot of hard work. Furthermore, verification of heterogeneous MASs\footnote{That is, MASs comprised of agents implemented in different APLs.} is not possible. Recently, others have proposed a way to verify heterogeneous MASs by translating programs into a common metalanguage, meta-APL \cite{Doan+2014}. However, since this approach requires a translation of the program, we need to convince ourselves that the translation is faithful to the original program. 

In this paper, we are focusing on another approach for verifying agent systems, which is based on an extended version of Java PathFinder (JPF) \cite{Visser+2003} called Agent JPF (AJPF), which takes advantage of the advanced model checking features of JPF, while making it possible to verify properties relevant to intelligent agents. AJPF can be used as-is for potentially any APL implemented in Java, but its real power shows, when combined with the \emph{agent infrastructure layer} (AIL). AIL is designed so that interpreters of semantically well-defined agent programming languages can be implemented using it \cite{Dennis+2011}, and has been optimized for model checking in AJPF, by using techniques such as state-space reduction. AIL comes with a simple APL, Gwendolen \cite{Dennis+2008}, which provides the default semantics for AIL. An AIL agent has a belief base, possibly a rule base, goals, plans and intentions. Furthermore, the agent has a reasoning cycle, which executes the implementation of the operational semantics. We will not go into details with all the different components of AIL, but refer to \cite{Dennis+2011} for a detailed description. 

\subsection{Specifying properties}

Model checking agent systems is only useful, if we can specify desirable properties in a language that can incorporate the mental attitudes of agents. In AJPF, these properties are specified in the \emph{property specification language} (PSL) \cite{Dennis+2011}. PSL is a linear-time temporal logic (LTL) with additional modal operators for beliefs, goals, intentions, actions and percepts. We can thus specify formulas that should hold in the system. 

The full PSL syntax is given below. \emph{ag} is the agent's name, $f$ is a ground first-order atomic formula.
$$
\phi ::= \textbf{B}(\mathit{ag}, f)~|~\textbf{G}(\mathit{ag}, f)~|~\textbf{A}(\mathit{ag}, f)~|~\textbf{I}(\mathit{ag}, f)~|~\textbf{P}(f)~|~\phi\lor\phi~|~\neg\phi~|~\phi\textsf{U}\phi~|~\phi\textsf{R}\phi
$$

$\textbf{B}(\mathit{ag}, f)$ is true if agent \emph{ag} believes $f$ to be true, $\textbf{G}(\mathit{ag}, f)$ is true if the agent has $f$ as a goal. $\textbf{A}$ represents actions, $\textbf{I}$ intentions and $\textbf{P}$ properties of the environment. The LTL formulas $\textsf{U}$ and $\textsf{R}$ represents ``until'' and ``release'', respectively. The temporal operators $\Diamond$ (eventually) and $\Box$ (always) can be derived from $\textsf{U}$ and $\textsf{R}$. 

The underlying semantics of the modal operators depends on the MAS being verified. When using \aorta\, we can specify the semantics of beliefs as follows:
$$
\mathit{MAS}\models\textbf{B}(\mathit{ag}, f) \quad\mathit{iff}\quad \mathit{MS}_{\mathit{ag}}\models \textsf{bel}(f),
$$
where \emph{MAS} is the multi-agent system, and $\mathit{MS}_{\mathit{ag}}$ is agent \emph{ag}'s mental state as defined in section \ref{sec:aorta}. The semantics of the other modal operators can be given in a similar way. 

\subsection{AJPF}\label{sec:ajpf}

Agent JPF is a module for the Java PathFinder \cite{Visser+2003}. JPF consists of an implementation of the Java Virtual Machine (JVM), which can execute all paths through a program in order to verify some predefined properties about the program. Since the state space often explodes, JPF employs state matching in order to reduce the number of states explored. 

AJPF implements a controller, which takes care of execution of each of the agents in the system. At each time step, it checks whether the system is in an end state\footnote{An end state is defined as a state where every agent is sleeping and the environment is not changing, thus nothing new can happen after such state has been reached.} and should terminate, and otherwise it decides which agent to execute. This decision is made by a \emph{scheduler}, which keeps a list of \emph{active} agents, i.e. agents that do not want to sleep. The model checker can then branch out at each of these states and execute each of the active agents (by choosing one path, executing it until reaching an end state and then backtracking).

Each agent in an APL (either an AIL-enabled or an existing APL) needs to implement the \texttt{MCAPLLanguageAgent} interface, which is used by AJPF to perform all the steps necessary for the verification of a system:
\begin{itemize}
  \item Perform a reasoning step (\texttt{MCAPLreason()}).
  \item Decide to put the agent to sleep (\texttt{MCAPLwantstosleep()}) or wake it up (\texttt{MCAPLwakeup()}).
  \item Check if a property holds (\texttt{MCAPLbelieves(fml)}, \texttt{MCAPLhasGoal(fml)}, etc.).
\end{itemize}

AJPF provides a listener, which is used to verify the properties specified in PSL. This can be done by first building a B\"{u}chi automaton that represents the property, and then compute the global behavior of the system by executing it. The product of these, the product automaton, can then be used to check if the property is violated. A property is violated if there exists a path to an accepting state \cite{Courcoubetis+1992}. The implementation of the model checker in AJPF employs techniques that allows progressively building the product automaton \cite{Gerth+1995}, making the verification process more efficient.

\section{Verification of Organizations in Multi-Agent Systems}\label{sec:mco}
In the previous section, we described the AJPF framework, which can be used for verification of MASs. We now turn to verification of organizational MASs. In \cite{Dignum+2011}, several desirable properties of organizational MASs were define: what makes an organization well-defined, good, effective, etc. These properties require not only a way to express the beliefs of the agents in the system, but also the state of the organization. For example, a good organization \emph{``is an organization such that if the organization has the capability to achieve $\phi$ and there is a group of roles in the organization responsible for realizing it, then the roles being in charge have a chain of delegation to roles that are played by agents in $A_i$ that are actually capable of achieving it''} \cite{Dignum+2011}. Being able to verify that a system satisfies these properties would be a large step towards convincing oneself that the system actually works. Verification of organizational aspects has been investigated before \cite{Astefanoaei+2008,Dennis+2009,Huget+2002,Vigano+2007}, but usually only by considering the internals of each agent as a black box. Model checking of electronic institutions specified in the ISLANDER framework was explored in \cite{Huget+2002}. By translating an ISLANDER specification into MABLE, a language for automatic verification of MASs, the system can be verified using the SPIN model checker.

The work most similar to ours is described in \cite{Dennis+2009}, where a programming language for normative MASs is implemented in AIL. Agents in the system can interact with an organization, and the system can then verify various properties of both the agents and the organization. Our integration with AIL differs in that we only verify properties of the agents, but these properties may include organizational properties, as defined in the \aorta\ component. Since \aorta\ is not tightly coupled to a specific APL, the integration with AIL allows us to perform verification of existing agents with additional properties concerning an organization.

In the remainder of this section, we show how \aorta\ can be integrated into AIL, such that 1) interpreters implemented in AIL can make use of \aorta, and 2) verification of these systems is possible using AJPF\footnote{\aorta\ and the integration with AIL is open source and is available at \url{http://www2.compute.dtu.dk/~ascje/AORTA/}}. 

\subsection{Specifying organizational properties}
In order to verify properties about organizational beliefs and options, we need to be able to express the properties in PSL. We therefore extend the PSL syntax to incorporate such properties:
$$
\psi ::= \phi~|~\textbf{Org}(\mathit{ag}, f)~|~\textbf{Opt}(\mathit{ag}, f)
$$
The interpretation of $\textbf{Org}(\mathit{ag}, f)$ is given as:
$$
\mathit{MAS} \models \textbf{Org}(\mathit{ag}, f) \quad\mathrm{iff}\quad \mathit{MS}_{\mathit{ag}} \models \textsf{org}(f),
$$
where \emph{MAS} is the multi-agent system (AIL+\aorta) and $\emph{MS}_{\mathit{ag}}$ is agent \emph{ag}'s mental state as defined in section \ref{sec:aorta}. Similarly, the interpretation of $\textbf{Opt}(\mathit{ag}, f)$ is:
$$
\mathit{MAS} \models \textbf{Opt}(\mathit{ag}, f) \quad\mathrm{iff}\quad \mathit{MS}_{\mathit{ag}} \models \textsf{opt}(f).
$$

In AJPF, we have implemented the extended PSL by adding functions checking whether an agent has organizational beliefs or options to the \texttt{MCAPLLanguage\-Agent} interface, which defines the methods needed by AJPF to perform model checking. 

\subsection{Verifying AORTA}
Verification of the agents with an \aorta\ component requires (1) an integration of the \aorta\ architecture within existing AIL agents, and (2) the ability to verify properties about organizational beliefs and organizational options. 

We have integrated the \aorta\ architecture in AIL, allowing existing interpreters implemented in AIL to take advantage of the \aorta\ organizational reasoning component. The \texttt{AortaAILAgent} extends the \texttt{AILAgent} class as follows:
\begin{description}
  \item[\texttt{MCAPLreason(int flag)}] Executes the \aorta\ reasoning cycle before calling the AIL agent's own reasoning cycle.
  \item[\texttt{MCAPLhasOrganizationalBelief(MCAPLFormula phi)}] Returns true if $\Sigma_o$ contains \texttt{phi}.
  \item[\texttt{MCAPLhasOrganizationalOption(MCAPLFormula phi)}] Returns true if $\Gamma_o$ contains \texttt{phi}.
  \item[\texttt{wantstosleep()}] Returns true if \emph{both} \aorta\ and the AIL agent wants to sleep. \aorta\ wants to sleep if the last execution did not change anything.
  \item[\texttt{addBel(...)}/\texttt{addGoal(...)}/\texttt{delBel(...)}/\texttt{removeGoal(...)}] Responsible for synchronization of knowledge bases.
  \item[\texttt{newMessages(Set<Message> msgs)}] Checks if any of the incoming messages are \emph{organizational messages} and if so, lets \aorta\ handle them. Otherwise, they are forwarded to the AIL agent.
\end{description}

We have furthermore implemented an \texttt{AILBridge}, which is responsible for updating the AIL agent, when \aorta\ performs actions that change the belief base or goal base.

\section{Evaluation of AIL+\aorta}\label{sec:eval}
In this section, we evaluate our integration of \aorta\ in AIL. As we shall see, the example is small enough to generate the entire state space within reasonble time (takes approximately 10 minutes), so we evaluate the system in two ways: 
\begin{enumerate}
  \item We generate the product automata on the fly by executing the agent system, while verifying each of the properties.
  \item We first generate the entire state space for the system, and use it to verify each of the properties.
\end{enumerate}
The first method is practical for large (possible infinite) systems, or properties that can be verified quickly (e.g. that the agents eventually enacts a role, since this is the first thing happening in our system). The second method is practical for verifying many properties in a single, finite system, since the time used for verification of each property is significantly lower than the time spend generating the state space. 

\subsection{Example: Writing a paper}
\begin{lstlisting}[float,caption={Gwendolen-program for writing a paper.},label={lst:masprogram}]
GWENDOLEN
:name: alice
:Initial Beliefs:
:Belief Rules:
:Initial Goals:
editor [achieve]
:Plans:
+!editor [achieve] : {True} <- +editor;
+!wtitle [achieve] : {True} <- +wtitle;
+!wabs [achieve] : {True} <- +wabs;
+!wsectitle [achieve] : {True} <- +wsectitle;
+!fdv [achieve] : {True} <- +fdv;
+!wcon [achieve] : {True} <- +wcon;
+!sv [achieve] : {True} <- +sv;

:name: bob
:Initial Beliefs:
:Belief Rules:
:Initial Goals:
writer [achieve]
:Plans:
+!writer [achieve] : {True} <- +writer;
+!wsec [achieve] : {True} <- +wsec;
+!wref [achieve] : {True} <- +wref;
\end{lstlisting}

To illustrate the capabilities of the model checker for AIL+\aorta, we use a simple example of a group of agents aiming to write a scientific paper using an organizational specification to help them collaborate (inspired by \cite{Hubner+2010}). In the example, an editor should create a first draft version (\emph{fdv}), consisting of a title (\emph{wtitle}) and an abstract (\emph{wabs}) and the section titles (\emph{wsectitle}). The submission version (\emph{sv}) is then created by letting a number of writers write the sections (\emph{wsec}) and the references (\emph{wref}), while the editor writes the conclusion (\emph{wconc}). The writers depend on the editor for the completion of \emph{fdv}, while the editor depends on the writers for the completion of \emph{wsec} and \emph{wref}. 

We consider two agents: \emph{Alice}, capable of editing the paper and \emph{Bob}, capable of being a writer. We have implemented the agents in Gwendolen. The implementation of the agents is shown in listing \ref{lst:masprogram}. Since the focus is not on verification of agents writing a paper, but rather on the organizational coordination mechanisms of \aorta, the implementation of each objective is very simple. Note that the initial goals of each agent are artificial goals, which are used to generate a role enactment option, since \aorta\ currently only supports generating role enactment options based on the agent's goals. 

\begin{lstlisting}[float,caption={\aorta-program for writing a paper.},label={lst:aortaprogram}]
role(R) : true => enact(R).
obj(bel(O)) : bel(me(Me)), org(obl(Me,_,bel(O),_)) => commit(O).
send(_, tell, org(rea(Me,R))) 
  :  bel(me(Me), agent(Ag), Ag\=Me), ~(bel(sent(Ag, org(rea(Me,R))))) 
  => send(Ag, org(rea(Me,R))).
send(R, achieve, O) 
  :  org(rea(Ag, R)), bel(me(Me), Ag\=Me), ~(bel(sent(Ag, goal(O)))) 
  => send(Ag, goal(O)).
send(R, tell, O) 
  :  org(rea(Ag, R)), bel(me(Me), Ag\=Me), ~(bel(sent(Ag, bel(O))))
  => send(Ag, bel(O)).
\end{lstlisting}

The agents are enriched with an \aorta\ component that enables them to enact roles, commit to objectives and coordinate. The \aorta-program for the agents is shown in listing \ref{lst:aortaprogram}. The agents are able to delegate goals (the \texttt{send(R, achieve, O)} option) and inform about completion of goals (the \texttt{send(R, tell, O)} option). They commit to objectives that they are obliged to complete, and for simplicity simply enact a role, if it is considered an option.

\subsection{Main results}

Our system was evaluated\footnote{We evaluated the system using Java 7 on a laptop with a dual core 2.80 GHz Intel i7 CPU and 8 GB RAM running Windows 8.1.} using the properties listed in table \ref{tab:properties}. Our results are divided into two sets: 1) the on the fly verification, for which we specify the number of states explored and the time used, and 2) the complete state space verification, in which we only state the time used, since in that case, the number of states is constant (the example contains 251 states).

Properties \ref{rea_a} and \ref{rea_b} check that the agents eventually enacts their roles. Property \ref{error} tries to verify that \emph{Alice} eventually enacts the \emph{writer} role, but fails, since that is not the case. Properties \ref{rea_tell_ab} and \ref{rea_tell_ba} check that the agents furthermore eventually knows about the other agent's enactment. Property \ref{one_obl} verifies that whenever \emph{Alice} is obliged to achieve \emph{wabs}, she will eventually believe that she has achieved it. Property \ref{one_dep} verifies that when \emph{Alice} believes she has completed \emph{fdv}, she will eventually inform \emph{Bob}, because of the dependency relation between their roles. Finally, property \ref{paper_written} verifies that the paper is eventually written. Note that the time difference between properties \ref{rea_a} and \ref{rea_b} are due to the way the AJPF scheduler executes the agents: \emph{Bob} is only executed once AJPF has detected that \emph{Alice} has nothing more to do.

Since the time complexity of LTL model checking is exponential in the length of the formula \cite{Baier+2008}, we furthermore verified larger formulas to get an assessment of the model checker's capabilities. Property \ref{all_rea} verifies that the agents know their own role and the role of the other agent. Property \ref{all_obl} verifies that for every obligation it is the case that it is eventually satisfied. Property \ref{all_dep} verifies that all dependency relations are used for generating options, and finally, \ref{all} verifies all of the properties above.

\renewcommand{\arraystretch}{1.1}
\setlength{\tabcolsep}{6pt}
\begin{table}[t]
\newcounter{properties}
\newcommand{\rprop}[1]{\refstepcounter{properties}\label{#1}\theproperties}
\centering
\caption{The properties that were verified by AJPF. The results are specified for 1) the on the fly verification and 2) the complete state space verification.}\label{tab:properties}
\scriptsize
\begin{tabular}{l|l||r|r|r}
 & \textbf{Property} & \textbf{States}$^1$ & \textbf{Time}$^1$ & \textbf{Time}$^2$  \\
\hline
\rprop{rea_a} & $\Diamond\textbf{Org}(\texttt{alice}, \mathit{rea}(\mathit{alice}, \mathit{editor}))$ & 6 & 0:14 & $<$10ms \\
\rprop{rea_b} & $\Diamond\textbf{Org}(\texttt{bob}, \mathit{rea}(\mathit{bob}, \mathit{writer}))$ & 6 & 0:26 & $<$10ms \\
\rprop{error} & $\Diamond\textbf{Org}(\texttt{alice}, \mathit{rea}(\mathit{alice}, \mathit{writer}))$ & 11 & 0:40 & $<$10ms \\
\rprop{rea_tell_ab} & $\Diamond\textbf{Org}(\texttt{alice}, \mathit{rea}(\mathit{bob}, \mathit{writer}))$ & 21 & 1:38 & 15ms \\
\rprop{rea_tell_ba} & $\Diamond\textbf{Org}(\texttt{bob}, \mathit{rea}(\mathit{alice}, \mathit{editor}))$ & 27 & 1:48 & 15ms \\
\rprop{all_rea} & 
  (\ref{rea_a}) $\land$ (\ref{rea_b}) $\land$ (\ref{rea_tell_ab}) $\land$ (\ref{rea_tell_ba})
  & 31 & 2:24 & 85ms \\
\rprop{one_obl} & 
	\makecell[l]{$\Box(\textbf{Org}(\texttt{alice}, \mathit{obl}(\mathit{alice}, \mathit{editor}, \mathit{wabs}, \mathit{fdv}))$ \\
	\qquad$\rightarrow \Diamond\textbf{B}(\texttt{alice}, \mathit{wabs}))$} 
	& 251 & 8:46 & 110ms \\
\rprop{all_obl} & 
  \makecell[l]{$\Box( \bigwedge \textbf{Org}(\texttt{ag}, \mathit{obl}(\mathit{ag}, \mathit{role}, \mathit{obj}, \mathit{deadline}))$ \\
  \qquad$\rightarrow \Diamond\textbf{B}(\mathit{ag}, \mathit{obj}))$}
  & 251 & 10:33 & 2740ms \\
\rprop{one_dep} & 
  \makecell[l]{$\Box(\textbf{Org}(\texttt{alice}, \mathit{dep}(\mathit{writer}, \mathit{editor}, \mathit{fdv})) \land \textbf{B}(\texttt{alice}, \mathit{fdv})$ \\
  \qquad$\rightarrow \Diamond\textbf{B}(\texttt{alice}, \mathit{sent}(\mathit{bob}, \mathit{bel}(\mathit{fdv}))))$} 
  & 251 & 8:51 & 25ms \\
\rprop{all_dep} & 
  \makecell[l]{$\Box \bigwedge \textbf{Org}(\texttt{ag}, \mathit{dep}(\mathit{role1}, \mathit{role2}, \mathit{obj})) \land \textbf{B}(\texttt{ag}, \mathit{obj})$ \\
  \qquad$\rightarrow \Diamond\textbf{B}(\mathit{ag}, \mathit{sent}(\mathit{ag}_\mathit{rea(role1)}, \mathit{bel(obj)}))$}
  & 251 & 9:15 & 95ms \\
\rprop{paper_written} & $\Diamond\textbf{B}(\texttt{alice}, \mathit{sv})$ & 167 & 8:34 & $<$10ms \\
\rprop{all} & 
  (\ref{all_rea}) $\land$ (\ref{all_obl}) $\land$ (\ref{all_dep}) $\land$ (\ref{paper_written})
  & 251 & 13:10 & 15812ms \\
\end{tabular}
\end{table}

As expected, once the state space has been fully generated, verification of each property is quite fast. We also see that verification of larger formulas takes much more time (e.g. property \ref{all_obl} compared to property \ref{one_obl}), which is expected, since the formulas are LTL formulas. Obviously, generating the entire model beforehand is only necessary, when we want to validate several properties, since in several of the cases, the entire state space does not need to be explored. 

Our experiments have further shown that one execution of an \aorta\ cycle is on average 10 times slower when executed using JPF compared to executing the system on the host JVM (i.e. without running the verification process). Even though some decrease in performance is expected, it should be possible to improve this by implementing the operational semantics of \aorta\ using AIL. In that case, we have to convince ourselves that the semantics are correctly implemented in AIL (i.e., the functionality should correspond to the functionality of the existing implementation). Still, the advantage of verifying the existing implementation is that the results are directly applicable (i.e., we know that the agents will inform each other about their roles, that they will conform with their obligations, etc.).

In section 4, we mentioned that an organization could be considered \emph{good} if the roles are related in such a way that the objectives of the organization will be delegated to the agents that can actually achieve them. We cannot directly specify general properties like this in PSL, so instead, they must be specified using the specific properties relevant to the given system. In our case, we can specify this using the dependency relations (property \ref{all_dep}) and the fact that the paper is eventually written (property \ref{paper_written}). Using the organizational model and the agent programs it should be possible to generate such specifications, but that is out of scope for this paper.

\section{Conclusion}\label{sec:conc}

As agent systems gain popularity and become increasingly complex, the possibility to understand every part of a system becomes difficult. By model checking agent systems, it will be possible to verify that the agents of the system behave as expected and that the outcome is satisfactory. We have discussed some of the work done concerning model checking of agent programming languages, focusing especially on the generic Agent Java PathFinder, which enables model checking of potentially any kind of agent programming language. Furthermore, by allowing the implementation of an APL interpreter \emph{within} AJPF (using the Agent Infrastructure Layer), it is possible to optimize the model checking process, making it feasible for larger systems. 

In increasingly complex systems, there is often a need for regulation, since agents may come from different sources and cannot as such be forced to perform actions required to achieve the system objectives. This further creates a need for model checking, since the complexity again increases. In this paper, we have shown that our framework for organizational reasoning, \aorta, can be model checked using an extended version of AJPF. We have integrated \aorta\ in AIL and have verified properties about a system implemented in an APL using \aorta\ for organizational reasoning. We have verified that the agents of the system enacts roles, coordinate role enactment and that they can successfully delegate tasks using an organizational model describing roles and their relations.

The \aorta\ framework is not tightly integrated with AIL, but is rather used as a library, which means that many of the optimization techniques of AJPF cannot be used. Furthermore, a single step of the \aorta\ reasoning cycle is considered atomic in the current implementation, making it impossible to verify properties about the internals of \aorta. Even though the system, in its current state, makes it possible to verify interesting properties about a system, it would be interesting to address these shortcomings in the future. However, as we have mentioned, even though the advantages of implementing the operational semantics of \aorta\ in AIL rather than using it as a library are desirable, they should be weighed against the drawback of not verifying the actual system, but a (hopefully) equivalent one (in terms of operational semantics).  We believe that both approaches have their merits, and an implementation of the operational semantics in AIL will be also a useful contribution to model checking of organization-aware agents. An obvious direction for future work is thus to implement the operational semantics in AIL. 

Finally, even though the example used in this paper shows that it is possible to verify properties about organization-aware agents, it is rather small and it would be interesting to verify larger, more complex systems containing more than a handful of agents.

\bibliographystyle{splncs03}
\bibliography{bibliography}

\begin{thebibliography}{10}
\providecommand{\url}[1]{\texttt{#1}}
\providecommand{\urlprefix}{URL }

\bibitem{Astefanoaei+2008}
A{\c s}tef{\u a}noaei, L., Dastani, M., Meyer, J.J., de~Boer, F.S.: {A
  verification framework for normative multi-agent systems}. In: Intelligent
  and Multi-Agent Systems. pp. 54--65 (2008)

\bibitem{Baier+2008}
Baier, C., Katoen, J.P.: Principles of Model Checking (Representation and Mind
  Series). The MIT Press (2008)

\bibitem{Boissier+2013}
Boissier, O., van Riemsdijk, M.B.: {Organisational Reasoning Agents}. Agreement
  Technologies pp. 309--320 (2013)

\bibitem{Bordini+2008}
Bordini, R.H., Dennis, L.A., Farwer, B., Fisher, M.: {Automated Verification of
  Multi-Agent Programs}. In: 2008 23rd IEEE/ACM International Conference on
  Automated Software Engineering. pp. 69--78. IEEE (2008)

\bibitem{Bordini+2004a}
Bordini, R.H., Fisher, M., Visser, W., Wooldridge, M.: {Verifiable multi-agent
  programs}. Programming Multi-Agent Systems  LNCS(LNAI 3067),  72--89 (2004)

\bibitem{Bordini+2006}
Bordini, R.H., Fisher, M., Visser, W., Wooldridge, M.: {Verifying Multi-agent
  Programs by Model Checking}. Autonomous Agents and Multi-Agent Systems
  12(2),  239--256 (2006)

\bibitem{Bordini+2004b}
Bordini, R.H., Fisher, M., Wooldridge, M., Visser, W.: {Model checking rational
  agents}. IEEE Intelligent Systems  (2004)

\bibitem{Courcoubetis+1992}
Courcoubetis, C., Vardi, M., Wolper, P., Yannakakis, M.: Memory-efficient
  algorithms for the verification of temporal properties. In: Formal Methods in
  System Design. pp. 275--288 (1992)

\bibitem{Dennis+2009}
Dennis, L., Tinnemeier, N., Meyer, J.J.: {Model checking normative agent
  organisations}. In: Computational Logic in Multi-Agent Systems (2009)

\bibitem{Dennis+2008}
Dennis, L.A., Farwer, B.: {Gwendolen: A BDI Language for Verifiable Agents}.
  In: L{\"o}we, B. (ed.) Logic and the Simulation of Interaction and Reasoning.
  AISB, Aberdeen (2008), {AISB'08 Workshop}

\bibitem{Dennis+2011}
Dennis, L.A., Fisher, M., Webster, M.P., Bordini, R.H.: {Model Checking Agent
  Programming Languages}. Automated Software Engineering  19(1),  5--63 (2011)

\bibitem{Dignum+2011}
Dignum, V., Dignum, F.: A logic of agent organizations. Logic Journal of IGPL
  pp. 283--316 (2011)

\bibitem{Doan+2014}
Doan, T.T., Yao, Y., Alechina, N., Logan, B.: {Verifying heterogeneous
  multi-agent programs}. In: Autonomous Agents and Multi-Agent Systems. pp.
  149--156 (2014)

\bibitem{Gerth+1995}
Gerth, R., Peled, D., Vardi, M.Y., Wolper, P.: Simple on-the-fly automatic
  verification of linear temporal logic. In: Protocol Specification, Testing
  and Verification. pp. 3--18. Chapman \& Hall, Ltd. (1996)

\bibitem{Hubner+2010}
H\"{u}bner, J.F., Boissier, O., Kitio, R., Ricci, A.: Instrumenting multi-agent
  organisations with organisational artifacts and agents. Autonomous Agents and
  Multi-Agent Systems  20(3),  369--400 (2010)

\bibitem{Hubner+2002}
H\"{u}bner, J.F., Sichman, J.S., Boissier, O.: A model for the structural,
  functional, and deontic specification of organizations in multiagent systems.
  In: SBIA '02 Proceedings. pp. 118--128 (2002)

\bibitem{Huget+2002}
Huget, M.P., Esteva, M., Phelps, S., Sierra, C., Wooldridge, M.: {Model
  checking electronic institutions}. In: MoChArt 2002. pp. 51--58 (2002)

\bibitem{Jensen+2014a}
Jensen, A.S., Dignum, V.: {AORTA: Adding Organizational Reasoning to Agents}.
  In: Proc. AAMAS '14. pp. 1493--1494 (2014)

\bibitem{Jensen+2014b}
Jensen, A.S., Dignum, V., Villadsen, J.: The {AORTA} architecture: Integrating
  organizational reasoning in \emph{Jason}. In: 2nd International Workshop on
  Engineering Multi-Agent Systems (EMAS 2014). pp. 112--128 (2014)

\bibitem{Rao+1996}
Rao, A.S.: {AgentSpeak (L): BDI agents speak out in a logical computable
  language}. Agents Breaking Away (L) (1996)

\bibitem{Vigano+2007}
Vigan\`{o}, F.: {A framework for model checking institutions}. In: Model
  Checking and Artificial Intelligence. pp. 129--145 (2007)

\bibitem{Visser+2003}
Visser, W., Havelund, K., Brat, G., Park, S., Lerda, F.: {Model checking
  programs}. Automated Software Engineering  10,  203--232 (2003)

\end{thebibliography}

\end{document}